\documentclass[conference]{IEEEtran}
\IEEEoverridecommandlockouts
\usepackage{cite}
\usepackage{amsmath,amssymb,amsfonts}
\usepackage{algorithmic}
\usepackage{graphicx}
\usepackage{textcomp}
\usepackage{xcolor}
\def\BibTeX{{\rm B\kern-.05em{\sc i\kern-.025em b}\kern-.08em
    T\kern-.1667em\lower.7ex\hbox{E}\kern-.125emX}}
\begin{document}

\title{Comparative Review of Cloud Computing Platforms for Data Science Workflows\\

}

\author{\IEEEauthorblockN{Mohammad Rehman}
\IEEEauthorblockA{\textit{School of Computer Science and Applied Mathematics} \\
\textit{University of the Witwatersrand}\\
Johannesburg, South Africa \\
707386@students.wits.ac.za}
\and
\IEEEauthorblockN{Hairong Wang}
\IEEEauthorblockA{\textit{School of Computer Science and Applied Mathematics} \\
\textit{University of the Witwatersrand}\\
Johannesburg, South Africa \\
hairongwng@gmail.com}}

\maketitle

\begin{abstract}
With the advantages that cloud computing offers in terms of platform as a service, software as a service, and infrastructure as a service, data engineers and data scientists are able to leverage cloud computing for their ETL/ELT (extract, transform and load) and ML (machine learning) requirements and deployments. The proposed framework for the comparative review of cloud computing platforms for data science workflows uses an amalgamation of the analytical hierarchy process, Saaty’s fundamental scale of absolute numbers, and a selection of relevant evaluation criteria (namely: automation, error handling, fault tolerance, performance quality, unit testing, data encryption, monitoring, role based access, security, availability, ease of use, integration and interoperability). The framework enables users to evaluate criteria pertaining to cloud platforms for data science workflows, and additionally is able to recommend which cloud platform would be suitable for the user based on the relative importance of the above criteria. Evaluations of the criteria are shown to be consistent and thus the weighting of criteria against the goal of cloud service provider or cloud platform selection are sensible. The proposed framework is robust enough to accommodate for changes in criteria and alternatives, depending on user cloud platform requirements and scope of cloud platform selection.
\end{abstract}

\begin{IEEEkeywords}
Data science workflows, analytical hierarchy process (AHP), evaluation of criteria, cloud platform, cloud services
\end{IEEEkeywords}

\section{Introduction}
\noindent With an increase in migration to cloud-based extract, transform and load (ETL) operations, analytics, machine learning (ML) and reporting solutions, data scientists and analysts need to become comfortable with moving away from tools such as Microsoft Excel and on-premises databases for their data storage and compute requirements, and gain confidence in storage and compute technology offered by cloud service providers. Although there are the giants among the cloud computing service providers (CSPs), for example Microsoft\footnote{https://portal.azure.com}, Amazon\footnote{https://aws.amazon.com/console} and Google\footnote{https://console.cloud.google.com}, consumers can still opt to partner with many other CSPs such as Alibaba\footnote{https://home-intl.console.aliyun.com}, Oracle\footnote{https://cloud.oracle.com} and IBM\footnote{https://cloud.ibm.com}. \\
\indent As the migration to cloud platforms is a journey that could span many months to years, consumers need to be sure that the services offered by their desired CSP are a good fit for their requirements, and thus need to know the advantages and disadvantages of the different CSPs based on factors such as compute, security, governance, scalability, pricing, availability, etc. Therefore, this research has  investigated, evaluated and reviewed the ETL capabilities and the ML service integration of the cloud platforms offered by three popular cloud service providers: Microsoft Azure, Amazon Web Services, and Google Cloud Platform. The analytical hierarchy process was used to provide a framework for evaluating criteria pertaining to cloud based platforms specifically for data science workflows, for the use case of cloud service provider selection. The research shows that consumers can use multi-criteria decision making methodologies to determine which cloud provider would be most beneficial for them based on the relative importance of their requirements.\\
\indent The following sections in this report delve into the available literature, followed by an overview of the methodology used for experimentation and evaluation, and subsequently elaborate on the results obtained from the experiments while providing a concise discussion on the implications of the results. Finally, a conclusion is provided in terms of the overall outcomes of the research.

\section{Related Work}
\noindent In as early as 2008, Vouk had considered the use of cloud computing infrastructure for building out effective workflows \cite{b1}.The importance and need of on-demand as well as reserve-based provisioning of computational services, automation and security was highlighted as a measure of effectiveness, and this is true to-date. Integration of services, appropriate governance and security, and automation are all critical to having efficient and effective workflows within any cloud computing platform. 

\indent C. Chen, J. Chen, J. Liu and Y. Wen have presented the need for workflow scheduling on cloud platforms and how it can be implemented \cite{b2}, however, this was based on using external workflow scheduling platforms. Zhou et al. have investigated this further by incorporating parallelism into cloud-based workflows \cite{b3} and have shown that external workflow engines can be used (together with a run-time service) to leverage parallelism in order to achieve improvements when scaling workflows. 

\indent The generally accepted criteria for evaluating cloud platforms and services are: security, performance, accessibility and useability, scalability, and adaptability \cite{b4} which can be further explored thanks to the consolidation of past evaluations by Abdel-Basset, Mohamed and Chang \cite{b4}. Reference \cite{b4} has additionally provided individuals with a "neutrosophic multi-criteria decision analysis" methodology to determine the quality of service that a cloud provider offers. A methodology that has been used in many frameworks that aim to evaluate various aspects of cloud computing is the analytical hierarchy process, which aims to decompose multi-criteria decision problems into hierarchical segments for evaluation (both qualitatively and quantitatively) \cite{b5} \cite{b6} \cite{b7}. For security specifically, Ghazizadeh and Cusack have proposed a "Trust Framework" for evaluation of cloud computing technologies \cite{b8}. Additionally, the National Institute of Standards and Technology (U.S. Department of Commerce) have developed the "Cloud Usability Framework" that enables users to be informed of what cloud service providers need to adhere to and ensure is available within their platforms and services \cite{b9}. 

\indent Costa, Santos and Mira da Silva have looked into a significant number of criteria for the evaluation of cloud services in general, broadly within the realm of accountability, agility, management, performance, security and usability \cite{b10}. Metrics for evaluating cloud-based machine learning services were proposed and reviewed in 2017 \cite{b11}. The paper included, but was not limited to, metrics such as availability, data capacity, compatibility, documentation, functionality, integration, performance, pricing and usability. In a recent study looking at the selection of cloud services adequate for big data, the proposed important criteria were identified as: architecture, cost, functionality, performance, usability and vendor reputation \cite{b12}. In a quality of service evaluation approach study pertaining to managing cloud service evaluation and selection, a similar range of criteria were considered, namely: accountability, agility, assurance, cost, performance and security \cite{b13}. 

\section{Methodology}
\noindent A qualitative review of criteria pertaining to the top cloud computing platforms required an in-depth understanding of the various ETL processes available across the cloud platforms and thus necessitated the exploration of the services related to data science workflows prior to building out the workflows to gain familiarity with their functionality, governance and accessibility related capabilities, thus making these the criteria that were used within the evaluation framework. As such, each of the cloud computing platforms that were utilised explored the following capabilities, i.e. their sub-criteria:
\\
\\
\noindent\textbf{Functionality:} Automation, error handling, fault tolerance, performance quality and unit testing\\
\textbf{Governance} Data encryption, monitoring, security and role-based access\\
\textbf{Accessibility} Availability, ease of use, integration, interoperability
\\

\indent A focus was placed on services that enabled data engineering and
analytics in order to remain in scope of the proposed research. The
accessibility criteria were included in order to supplement the evaluation as a major driver of cloud adoption is the ability to learn and use the platform relatively easily to promote its adoption, as well as to have certainty that it is able to integrate well with applications and tools external to the platform, thus mitigating vendor lock-in. A similar workflow was developed using the various platforms and the experience gained while building out the end-to-end workflows was used as a basis for reviewing each of the criteria and sub-criteria from an entry level data scientist perspective, thus informing the valuation of each of the attributes within the framework.

\indent The analytical hierarchy process (AHP) was subsequently used to develop a framework \cite{b14} that evaluated the criteria pertaining to cloud platforms based on the valuations of the criteria and sub-criteria previously mentioned. In the evaluation process, Saaty's fundamental scale of absolute numbers \cite{b14} was used to attribute numerical values to linguistic associations of importance to a pair of relevant attributes. Although Microsoft Azure, Amazon Web Services and Google Cloud Platform were used as the cloud platforms in this research, the actual development of the framework used anonymised platform providers, i.e. CSP1, CSP2 and CSP3, as the intention of the research is to ultimately propose a feasible, dynamic evaluation framework to identify suitable cloud computing platforms for data science workflows rather than to compare Azure, AWS and GCP in an official and categorical manner.

\begin{figure}[htbp]
	\centering
	\includegraphics[width=0.9\linewidth]{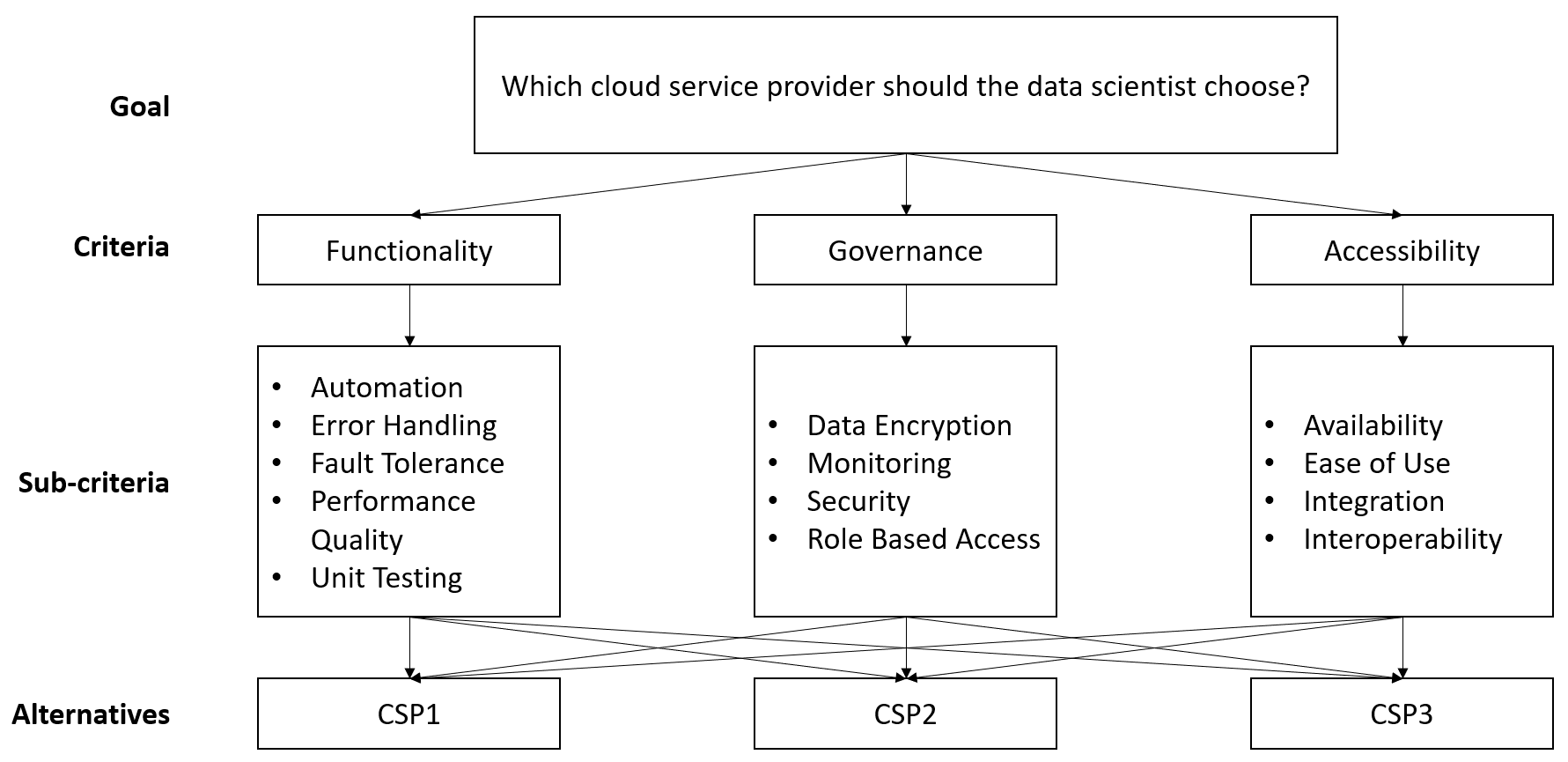}
	\caption{AHP model for determining cloud platform based on data science workflow requirements}
	\label{fig1}
\end{figure}

The context of this research is within the domain of a multi-criteria decision making problem, where a selection of alternatives need to be evaluated based on a set of criteria in order to decide on which alternative is the most favourable for a requirement. Therefore, the analytical hierarchy process can be used to develop a framework for decision making pertaining to the comparative review of a multitude of criteria that pertain to a set of cloud platforms for data science workflows, from the perspective of an entry level data scientist who has the goal of deciding which platform would be most suitable for the user's data science requirements. The hierarchy tree used for this research is illustrated in Fig.~\ref{fig1}

\indent The analytical hierarchy process was selected as it provides decision makers with a framework to attribute mathematical valuations to subjective opinionations
pertaining to a set of criteria in an overall evaluation, thereby enabling users to mathematically deduce which alternative is most suitable for the user against a defined goal. It does this by associating relative measures among a pair of criteria that are relevant to the user, thus propagating a pairwise matrix of relative importance for each of the
criteria. This can be developed further by means of using sub-criteria, which
provide a granular numeric representation of importance among the hierarchy of
attributes relevant to the user. As the process is based on subjective beliefs,
aggregated inputs from a myriad of users is beneficial, especially for important
decision making.\\

\section{Experimentation and Results}
\noindent The analytical hierarchy process was used to develop a framework for reviewing criteria pertaining to cloud computing platforms for data science workflows. The experimental process entailed the construction of ETL pipelines using the cloud-based ETL platforms of various (anonymised) cloud service providers to investigate attributes such as automation, error handling, fault tolerance, performance quality, unit testing, data encryption, monitoring, security, role based access, availability, ease of use, integration and interoperability. The attributes were subsequently used to review and evaluate the ETL platforms relative to one another based on the experience of an entry level data scientist, to determine which cloud platform would be the most suitable in terms of selection for carrying out data science workflows. In the evaluation process, Saaty's fundamental scale of absolute numbers was used to attribute numerical values to linguistic associations of importance to a pair of relevant attributes. This conversion of linguistic valuation to numeric valuation ensured that matrices containing the relative importance of criteria and sub-criteria were able to be constructed, which aided in assigning global weights to sub-criteria in terms of their overall importance. Subsequently, a similar approach made it possible to evaluate each of the cloud platforms relative to one another in order to determine the platform with the highest overall weighting in terms of the chosen sub-criteria, thus serving as a recommendation for the platform that would likely be most suitable for the data scientist.

\indent Once all criteria and sub-criteria evaluations were validated to be consistent, the weightings of the sub-criteria were converted to global weights. This was done by multiplying the weight of each sub-criteria with the weight of its parent criteria. Remembering that the sum of weights for criteria, as well as sub-criteria sets, was equal to $1$, the weights can be treated as percentages, thus the multiplication of sub-criteria weight and parent weight yields the weighting of a sub-criteria in terms of the goal of the analytical hierarchy process. As such, the sum of the global weights is also equal to $1$. Table~\ref{tab1} illustrates the global weights of each of the sub-criteria:
\begin{table}[htbp]
	\begin{center}
			\caption{Global sub-criteria weightings for the analytical hierarchy process}
		\begin{tabular}{ ccc } 
		    \hline
			\textbf{Criteria} & \textbf{Sub-criteria} & \textbf{Global Weight} \\
			\hline
			Functionality & Automation & 0.14 \\
			Functionality & Error Handling & 0.07 \\
			Functionality & Fault Tolerance & 0.02 \\
			Functionality & Performance Quality & 0.04 \\
			Functionality & Unit Testing & 0.28 \\
			Governance & Data Encryption & 0.02 \\
			Governance & Monitoring & 0.07 \\
			Governance & Security & 0.01 \\
			Governance & Role Based Access & 0.03 \\
			Accessibility & Availability & 0.02 \\
			Accessibility & Ease of Use & 0.17 \\
			Accessibility & Integration & 0.04 \\
			Accessibility & Interoperability & 0.08 \\
			\hline
		\end{tabular}
	\label{tab1}
	\end{center}
\end{table}

\indent Having obtained the weights of each alternative against each sub-criteria ultimately allowed for the calculation of the overall weighting of the alternatives against the goal, similar to how the sub-criteria were weighed against the goal, as shown in Table~\ref{tab1}. Specifically, this was done by multiplying the weight of each alternative per sub-criteria to the global weight of the sub-criteria. As a result, the final weighting of the alternatives were calculated and are demonstrated in Table~\ref{tab2}:
\begin{table}[htbp]
	\begin{center}
			\caption{Overall weightings of alternatives against the goal}
		\begin{tabular}{ ccc } 
			\hline
			& Weights & Implications \\ 
			\hline
			CSP1 & 0.28 & Suitable for user \\ 
			CSP2 & 0.19 & Least suitable for the user \\ 
			CSP3 & 0.53 & Most suitable for the user \\   
			\hline
		\end{tabular}
	\label{tab2}
	\end{center}
\end{table}

\indent The results indicate that based on the relative importance of criteria and sub-criteria to the user, the experience obtained by making use of the cloud platform provided by \textbf{CSP3} would be most suitable for the data scientist as it has the highest weighting based on the evaluation of the criteria and sub-criteria relevant to the tasks needed to be performed on said platform.

\section{Discussion}
\indent Within the scope of the research, the attributes selected were what the data scientist would look for when using the cloud platforms for data science workflows, thus ensuring that the criteria used to review and evaluate each platform were suitable. Additionally, as a comparative review was desired, and not a review to determine the absolute best platform, the analytical hierarchy process was perfect for achieving this. Pairwise comparisons were done between criteria and sub-scriteria to determine the numeric representation of the relative levels of importance among attributes. As the analytical hierarchy process has a means of evaluating consistency, the evaluations made when reviewing the attributes were able to be tested for any inconsistency and the outcomes for the evaluations pertaining to the criteria and sub-criteria were indeed consistent. Finally, the alternatives were reviewed against each of the sub-criteria in order to weigh the various alternatives (cloud platforms provided by the cloud service providers) against the goal. Ultimately, this allowed the framework to recommend a cloud provider to the user by answering the question present in the goal of the analytical hierarchy process with the alternative that yielded the highest weight, which in the experiment conducted for this research was CSP3. 

\indent Evaluation of multi-criteria systems can become quite complex,  moreso when the criteria are of a linguistic nature and are dependent on subjective beliefs. This was an additional motivation of using the analytical hierarchy process, along with Saaty's fundamental scale of absolute numbers, so that linguistic associations of relative importance between each of the attributes could be converted to numeric values upon which mathematical operations could be conducted to produce weightings of sub-criteria and alternatives against the goal. It enabled the framework to be both replicable and robust, as the analytical hierarchy process is a widely known methodology for multi-criteria decision analysis and thus can be followed to reproduce the experiments and results obtained in this report. Additionally, based on how the framework is set up, users can apply the framework to systems with varying criteria, sub-criteria and alternatives, depending on their requirements and accessibility to platforms.

\indent As such, having shown the consistency in evaluation as well as the ability for the framework to recommend a cloud service provider, the outcomes of the research efforts have been positive: a successful framework was developed using the analytical hierarchy process in order to review, compare and evaluate criteria pertaining to cloud platforms based on the required attributes, and to recommend a suitable platform for a user based on the review and subsequent evaluation of the attributes, while maintaining confidence in the evaluation process by means of determining that the evaluations of the criteria and sub-criteria were consistent.

\section{Conclusion}
\noindent Although a significant amount of research has been done regarding scheduling of data science workflows, cloud computing theory, and the comparison of cloud computing platforms in terms of concepts such as performance, pricing and storage, not much has been done with regards to evaluating or reviewing cloud based platforms for data science workflows specifically. Thus research was conducted to understand various data science workflow platforms, the requirements that a data scientist would have when using such platforms and the availability of existing frameworks or methodologies that would enable some one to conduct a comparative review of cloud platforms for data science workflows. Since no such framework was identified, a novel framework using the analytical hierarchy process, Saaty's fundamental scale of absolute numbers, and relevant attributes to a data science workflow was created. It was applied to a scenario where a data scientist needed to select a cloud platform that would be suitable for the user's requirements and the framework was successful in recommending a suitable cloud service provider to the data scientist. 

\indent The benefits of using the analytical hierarchy process were twofold: the first was that a measure, namely consistency, could be used to ensure that evaluations were done correctly, and the second was that it enabled the review and evaluations of criteria pertaining to the cloud platforms (alternatives) to be done in a relative manner depending on the experience of an entry level data scientist. In this way, decisive or categorical proposals of one cloud platform being objectively better than another could be avoided, with the additional benefit on users of the framework being able to have their own alternatives depending on which set of platforms would be feasible for them (depending on budget, vendor agreements, brand affiliation, etc.).

\indent Overall, a successful framework was created and demonstrated for the comparative review of cloud computing platforms for data science workflows. The evaluations were deemed consistent and the framework is dynamic enough to be tailored to user requirements (attributes can be changed as needed) and platform scope (relevant alternatives to the user's circumstances can be chosen).

\end{document}